%% file: main_dc.tex
\documentclass[a4paper,fleqn]{cas-dc}
\usepackage[numbers,sort,compress]{natbib}
\usepackage{adjustbox}
\usepackage[ruled]{algorithm2e}
\usepackage{placeins}

\def\tsc#1{\csdef{#1}{\textsc{\lowercase{#1}}\xspace}}
\tsc{WGM}
\tsc{QE}

\let\printorcid\relax 

\begin{document}
\let\WriteBookmarks\relax
\def\floatpagepagefraction{1}
\def\textpagefraction{.001}
\sloppy

\shorttitle{Improving COVID-19 CT Classification of CNNs by Learning  Parameter-Efficient Representation}    

\shortauthors{Y. Xu \textit{et~al}.}  

\title [mode = title]{Improving COVID-19 CT Classification of CNNs by Learning  Parameter-Efficient Representation}  

\author[1]{Yujia Xu}
\credit{Conceptualization, Methodology, Writing original draft}

\author[1]{Hak-Keung Lam}
\cormark[1]
\cortext[1]{Corresponding author}
\credit{Method improvement, Review \& editing, Supervision}

\author[1]{Guangyu Jia}
\credit{Method improvement}

\author[1]{Jian Jiang}
\credit{Review \& editing}

\author[1]{Junkai Liao}
\credit{Review \& editing}

\author[1]{Xinqi Bao}
\credit{Review \& editing}

\affiliation[1]{organization={Department of Engineering},
                addressline={King's College London, Strand}, 
                city={London},
                postcode={WC2R 2LS},
                country={United Kingdom}}
                
\nonumnote{Email addresses: \{yujia.xu, hak-keung.lam, guangyu.jia, jian.jiang, junkai.liao, xinqi.bao\}@kcl.ac.uk}

\input{sections/0_abstract}
\maketitle

\input{sections/1_introduction}
\input{sections/2_related_works}
\input{sections/3_method}
\input{sections/4_results}
\input{sections/5_ablation}
\input{sections/6_discussion}
\input{sections/7_conclusion}
\printcredits
\input{sections/acknowledgement}


\bibliographystyle{model1-num-names}
\bibliography{refs.bib}

\input{sections/appendix}


\end{document}

%% file: sections/0_abstract.tex
\begin{abstract}
COVID-19 pandemic continues to spread rapidly over the world and causes a tremendous crisis in global human health and the economy. Its early detection and diagnosis are crucial for controlling the further spread. Many deep learning-based methods have been proposed to assist clinicians in automatic COVID-19 diagnosis based on computed tomography imaging. However, challenges still remain, including low data diversity in existing datasets, and unsatisfied detection resulted from insufficient accuracy and sensitivity of deep learning models. To enhance the data diversity, we design augmentation techniques of incremental levels and apply them to the largest open-access benchmark dataset, COVIDx CT-2A. Meanwhile, similarity regularization (SR) derived from contrastive learning is proposed in this study to enable CNNs to learn more parameter-efficient representations, thus improving the accuracy and sensitivity of CNNs.
The results on seven commonly used CNNs demonstrate that CNN performance can be improved stably through applying the designed augmentation and SR techniques. In particular, DenseNet121 with SR achieves an average test accuracy of $99.44\%$ in three trials for three-category classification, including normal, non-COVID-19 pneumonia, and COVID-19 pneumonia. And the achieved precision, sensitivity, and specificity for the COVID-19 pneumonia category are $98.40\%$, $99.59\%$, and $99.50\%$, respectively. These statistics suggest that our method has surpassed the existing state-of-the-art methods on the COVIDx CT-2A dataset.
\end{abstract}


%% file: sections/1_introduction.tex
\section{Introduction}
\label{sec:introduction}
The Coronavirus Disease 2019 (COVID-19) has become a worldwide pandemic and infected over $493$ million people till April 2022 \cite{WHO_dashboard}. Its increasingly high infectivity and fatality rate due to strain variation are threatening human health and damaging the global economy \cite{mckibbin2020economic, iacobucci2021covid, Mahasen597covid}. The efficient reproductive number of the virus in many countries remains high, as reported in \cite{global_rvalue}, indicating COVID-19 continues spreading quickly around the world. Thereby, a timely and efficient diagnosis is crucial for the treatment of COVID-19 positive patients and the control of further disease spread.

In the early diagnosis of COVID-19 infection, real-time reverse transcription polymerase chain reaction (RT-PCR) is the primary choice due to its convenience and high specificity. However, research results \cite{kucirka2020variation, li2020stability, tahamtan2020real} have suggested that RT-PCR is not sensitive enough that some infected patients turned out to be positive even after several negative tests. These false-negative cases might continue to infect their close contacts without isolation or develop into severe illness. Chest computed tomography (CT) is a supplementary screening tool to RT-PCR since CT has higher sensitivity in detecting infection, indicated by institutes \cite{kovacs2021sensitivity, fang2020sensitivity, xie2020chest}. The high cost and hours of scanning time of CT are not affordable for all institutes. Thus CT can be more suitable in some scenarios where patients have suspicious negative RT-PCR tests, or patients are in need of timely diagnosis, or the RT-PCR test kits are undersupplied.


Since the pandemic started, researchers have been exploring the potential of convolutional neural networks (CNNs) in COVID-19 CT classification and reported high accuracy without clinician intervention. CNNs are a kind of deep learning technique dominating computer vision tasks.
For example, Gunraj \textit{et al}. \cite{Gunraj2020ct1} introduced a large-scale open-access COVID-19 CT dataset (COVIDx CT-1) and trained a COVID-19-specific tailored CNN on it. Panwar \textit{et al}. \cite{panwar2020deep} utilized transfer learning to inherit cross-domain knowledge to improve the model performance. All these researches reveal that CNNs have the potential to serve as an assistant to help clinicians in COVID CT diagnosis. 

Although CNNs have achieved remarkable results in CT diagnosis, challenges remain before they can be put into practical use. Deep learning methods often require large-scale standard datasets, while the existing COVID-19 CT datasets are insufficient. Also, CT scans collected from different institutes have inconsistent characteristics like orientation, brightness, etc. The trained models might be more sensitive to these irrelevant information rather than the pneumonic pathologies that really matter. Furthermore, the increasingly great capability of CNN-based models may not be fully fulfilled given the limited data sources. Hence, methods for learning more parameter-efficient representations are crucial for mitigating the data insufficiency issue and improving the classification performance.

By addressing these problems above, a more reliable COVID-19 CT classification system can reduce the workload of clinicians and provide more accurate and sensitive computer-aided diagnoses. Motivated by these factors, this study aims to use deep learning techniques to improve the COVID-19 CT classification performance of commonly used CNNs. Particularly, to alleviate the data insufficiency and enhance the data diversity, we design and apply augmentation of incremental levels on the currently largest COVID-19 CT benchmark dataset (COVIDx CT-2A) \cite{Gunraj2021ct2}. Meanwhile, to find the optimal selection of CNN architectures and augmentation combinations, we explore seven commonly used CNN architectures under seven augmentation settings. The CNNs include SqueezeNet1.1 \cite{iandola2016squeezenet}, MobileNetV2 \cite{sandler2018mobilenetv2}, DenseNet121 \cite{huang2017densenet}, ResNet-18/34/50 \cite{he2016resnet}, and InceptionV3 \cite{szegedy2016inception}. Meanwhile, contrastive learning is one promising self-supervised method for enabling deep learning models to learn more parameter-efficient features.
We propose the similarity regularization (SR) derived from contrastive learning to learn more parameter-efficient representations and improve CNN classification. The experimental results demonstrate that SR can improve the classification performance of CNNs stably and surpass conventional contrastive learning. Our main contributions are summarized as follows:
\begin{enumerate}[a)]
    \item{We investigate the impacts of augmentation and model selection in COVID-19 CT classification for three classes, including normal, non-COVID-19 pneumonia (NCP), and COVID-19 pneumonia (CP).} 
    \item{We propose SR as a regularization term for learning more parameter-efficient representations. Comparisons between seven models with or without SR are conducted. The experimental results demonstrate that SR can improve classification stably without extra introduced model parameters during the test interface.}
    \item{Our proposed model DenseNet121-SR achieves $99.44\%$ test accuracy, and $98.40\%$ precision, $99.59\%$ sensitivity and $99.50\%$ specificity for COVID-19 positive class, achieving the state-of-the-art.}
    \item{On other COVID-19 CT datasets, i.e., SARS-CoV2 and COVIDx CT-1, our DenseNet121-SR outperforms the existing methods in terms of efficiency and accuracy.}
    \item{We extend the study to seven classic natural datasets and find that our DenseNet121-SR is superior to the original DenseNet121 for all tasks, indicating that our method can be generalized to general classification problems.}
\end{enumerate}


%% file: sections/2_related_works.tex
\section{Related Works}
\label{sec:related_works}

\subsection{COVID-19 Related Researches}
\label{subsec:covid_related_work}

\begin{table*}[!ht]
    \centering
    \caption{Comparison of classification metrics between multiple deep learning methods in four datasets. The precision, sensitivity, and specificity metrics are for COVID-19 positive class only. (The decimal places are kept consistent as reported in the publications.)}
    \label{tab:comparison_sota}
    \begin{tabular*}{\linewidth}{@{\extracolsep{\fill}}llrcccc@{}}
    \toprule
    Dataset       & Method   & Params. (M)  & Accuracy (\%) & Precision (\%) & Sensitivity (\%) & Specificity (\%) \\
    \midrule
    \multirow{7}{*}{SARS-CoV-2 \cite{angelov2020sars-cov-2}} 
        & Alshazly \textit{et al}. \cite{alshazly2021explainable}  & 86.74   & $99.4$   & $99.6$   & $99.8$     & $99.6$     \\
        & Silva \textit{et al}. \cite{silva2020covid}              & 4.78    & $98.99$  & $99.20$  & $98.80$    & -          \\
        & Kundu \textit{et al}. \cite{kundu2021fuzzy}              & 132.86  & $98.93$  & $98.93$  & $98.93$    & $98.93$    \\
        & Jaiswal \textit{et al}. \cite{jaiswal2020classification} & -       & $96.25$  & $96.29$  & $96.29$    & $96.21$    \\
        & Panwar \textit{et al}. \cite{panwar2020deep}             & 20.55   & $94.04$  & $95.30$  & $94.04$    & $95.86$    \\
        & Jangam \textit{et al}. \cite{jangam2021stacked}          & 202.87  & $93.5$   & $89.91$  & $98$       & -          \\
        & Wang \textit{et al}. \cite{wang2020contrastive}          & -       & $90.83$  & $95.75$  & $85.89$    & -          \\
    \midrule
        
    \multirow{5}{*}{COVID-CT \cite{zhao2020covidct}}
        & Chen \textit{et al}. \cite{chen2021momentum}             & -       & $88.5$   & $89.9$   & $88.6$     & -          \\
        & He \textit{et al}. \cite{he2020sample}                   & 0.55    & $86$     & -        & -          & -          \\
        & Polsinelli \textit{et al}. \cite{polsinelli2020light}    & 1.26    & $85.03$  & $85.01$  & $87.55$    & $81.95$    \\
        & Jangam \textit{et al}. \cite{jangam2021stacked}          & 202.87  & $84.73$  & $78.15$  & $94.9$     & -          \\
        & Wang \textit{et al}. \cite{wang2020contrastive}          & -       & $78.69$  & $78.02$  & $79.71$    & -          \\
    \midrule
        
    \multirow{2}{*}{COVIDx CT-1 \cite{Gunraj2020ct1}}  
        & Gunraj \textit{et al}. \cite{Gunraj2020ct1}              & 1.40    & $99.1$   & $99.7$   & $97.3$     & $99.9$     \\
        & Ter-Sarkisov  \cite{ter2020covid}                        & 34.14   & $91.66$  & $90.80$  & $94.75$    & -          \\
    \midrule
        
    \multirow{3}{*}{COVIDx CT-2A \cite{Gunraj2021ct2}}
        & Zhao \textit{et al}. \cite{zhao2021deep}                 & 23.51   & $99.2$   & $98.5$   & $98.7$     & $99.5$     \\
        & Gunraj \textit{et al}. \cite{Gunraj2021ct2}              & 0.45    & $98.1$   & $97.2$   & $98.2$     & $98.8$     \\
        & Gunraj \textit{et al}. \cite{Gunraj2020ct1}              & 1.40    & $94.5$   & $90.2$   & $99.0$     & $95.7$     \\
    \bottomrule
    \end{tabular*}
\end{table*}

CNNs are increasingly improving the COVID-19 CT classification with advanced algorithms and enhanced datasets. Numerous CNN-based methods achieving high accuracy have been proposed, indicating the potential of CNNs in assisting practical diagnosis. Some representative methods on four benchmark datasets are listed in Table~\ref{tab:comparison_sota}.

In the COVID-19 CT classification, there exist no golden standard datasets so far. The four widely employed open-access datasets \cite{angelov2020sars-cov-2, zhao2020covidct, Gunraj2020ct1, Gunraj2021ct2} in Table~\ref{tab:comparison_sota} differ in many aspects, including patient/scan distribution, collection sources, dataset size, class numbers, labelling quality, etc. Particularly, COVID-CT \cite{zhao2020covidct} and SARS-CoV-2 \cite{angelov2020sars-cov-2} are two small binary-classification datasets containing $812$ and $2,482$ CT scans for COVID-19 positive and non-COVID classes, respectively. Gunraj \textit{et al}. released a larger dataset COVIDx CT-1 \cite{Gunraj2020ct1} consisting of $104, 009$ scans for normal, NCP and CP classes upon which the authors later built COVIDx CT-2 \cite{Gunraj2021ct2}. COVIDx CT-2 is the largest existing dataset containing $194,922$ CT scans, combined from multiple data sources. Generally, data-driven methods like CNNs depends heavily on dataset size. This can be drawn from the classification metrics in Table~\ref{tab:comparison_sota} that the methods trained on larger datasets can roughly achieve higher performance. To ensure both the data diversity and satisfactory results, our study employs COVIDx CT-2A \cite{Gunraj2021ct2} as our target dataset.

Despite the various datasets, many CNN-based methods have been developed to continuously boost the classification performance. In particular, researchers often use transfer learning \cite{jaiswal2020classification, panwar2020deep, jangam2021stacked, silva2020covid, kundu2021fuzzy, alshazly2021explainable, shaik2022transfer} or ensemble learning \cite{jangam2021stacked, silva2020covid, kundu2021fuzzy, shaik2022transfer, kundu2021covid} to overcome the data insufficiency in small-scale datasets like SARS-CoV-2 and COVID-CT. For example, Jaiswal \textit{et al}. \cite{jaiswal2020classification} and Panwar \textit{et al}. \cite{panwar2020deep} utilized transfer learning to pre-train the weights of VGG19 and DenseNet201 on ImageNet and then fine-tuned on SARS-CoV-2, achieving $96.25\%$ and $94.04\%$ accuracy, respectively. Besides, ensemble learning, merging the decisions from multiple models into a more balanced decision, has been widely applied in some works through different merging approaches like weighted sum \cite{jangam2021stacked}, voting \cite{silva2020covid}, and fuzzy rank-based fusion \cite{kundu2021fuzzy}. However, ensemble learning is rarely applied in large-scale datasets like COVIDx CT-1/2 \cite{Gunraj2020ct1, Gunraj2021ct2}. Specifically, COVID-Net CT-1/2 L \cite{Gunraj2020ct1, Gunraj2021ct2} are two light-tailored CNNs whose architectures are finely designed by automatic neural architecture searching. The two models are extremely parameter-efficient and achieved $94.5\%$ and $98.1\%$ accuracy with only $0.45$MB and $1.40$MB parameters, respectively. Another research \cite{zhao2021deep} employed ResNet-50x1 pre-trained on ImageNet-21k and fine-tuned on COVIDx CT-2A, achieving $99.2\%$ accuracy. 

Drawn from the reviewed research works above, deep learning models can achieve higher performance in COVID-19 CT classification through the approaches that: 1) train models over data of higher diversity; 2) with finely designed neural networks; 3) ensemble the decisions from multiple models; 4) inherit out-of-domain classification knowledge. Although models can benefit from these aspects, the expensive computational cost of neural architecture searching and large-scale pre-training, and long execution time caused by over-parameterization should be considered as well.

\subsection{Contrastive Learning}
\label{subsec:contrastive_learning}
In recent years, supervised deep learning models of increasing complexity and depth have shown great progress in many large-scale applications like ImageNet classification \cite{he2016resnet, szegedy2016inception}. However, directly applying these models to COVID-19 datasets of smaller scales might cause over-parameterization. It means that model capacities cannot be fully fulfilled and the extracted representations are not parameter-efficient. One promising approach to addressing the issue is contrastive learning.

In the deep learning field, it is widely recognized that the model performance depends on the quality of their learned representation. Contrastive learning, also known as contrastive self-supervised representation learning, is one framework aiming at learning efficient representations without human-specified labels. In general, the main idea of contrastive learning is to project inputs into an embedding space where the embedded vectors of similar samples are closer while dissimilar ones are apart. More formally, for visual tasks, a pair of views augmented from one image is considered as a positive pair while pairs of views from different images are considered as negative pairs. Hence, contrastive learning models aim to maximize the representation similarity between positive pairs and minimize that between negative pairs. In practical tasks, contrastive learning often pre-trains the front representation extractors of deep learning models in a self-supervised manner, and then fine-tunes the pre-trained weights in a conventional supervised manner.

The state-of-the-art contrastive learning frameworks include MoCo \cite{he2020mocov1, chen2020mocov2}, SimCLR \cite{chen2020simclr, chen2020simclrv2}, SimSiam \cite{chen2021Simsiam}, SwAV \cite{caron2020swav}, BYOL \cite{grill2020byol}, etc. These frameworks mainly differ in terms of loss function, representation projection, and negative pair formation \cite{chen2021Simsiam}. And the differences further determine their requirements on complexity of augmentation policies and batch size. Normally, in order to obtain satisfactory result, contrastive methods depends on a large batch size to cover enough negative pairs \cite{chen2020simclr, chen2020simclrv2, he2020mocov1, chen2020mocov2}. Among these models, BYOL, SwAV and SimSiam are the contrastive frameworks requiring no negative pairs. In ImageNet linear classification experiments \cite{chen2021Simsiam}, BYOL achieves relatively better performance. This explains that we select BYOL as the basic framework for SR calculation as in Section \ref{subsec:similarity_reg}.

The success of contrastive learning has emerged some applications in COVID-19 CT diagnosis \cite{li2021multi, he2020sample, chen2021momentum}. He \textit{et al}. \cite{he2020sample} employed a MoCo-like \cite{he2020mocov1} framework to enhance the CT scans representations extracted by DenseNet169 and fine-tuned the network, achieving $86\%$ accuracy on COVID-CT \cite{zhao2020covidct}. Similarly, Chen \textit{et al}. employed the MoCo-v2-like \cite{chen2020mocov2} framework on the same dataset and reached $88.5\%$ accuracy within six shots. Li \textit{et al}. \cite{li2021multi} put the contrastive loss as a regularization term and trained their CMT-CNN in an end-to-end manner, obtaining $93.46\%$ accuracy. These studies suggest contrastive learning can boost classification performance by learning more efficient representations. 

%% file: sections/3_method.tex
\section{Method}
\label{sec:method}

\subsection{Augmentation of Incremental Levels}
\label{subsec:augmentation}
Data augmentation is vital for improving the performance of deep learning models, especially for contrastive learning \cite{chen2020simclr, chen2020mocov2}. However, the optimal selection for COVID-19 CT augmentation has not been studied. Inspired by the literature in Section \ref{sec:related_works}, we design and evaluate a series of augmentation operations of incremental levels as follows where "+" denotes the appended augmentation based on the previous level: \textbf{Level 0}: No augmentation; \textbf{Level 1} + RandomResizedCrop: Randomly obtain an image crop of size in the range $[0.08, 1]$ of the original size $256\times256$, and then randomly scale the crop according to an aspect ratio in the range $[3/4, 4/3]$. The scaled crop is finally resized to the original size; \textbf{Level 2} + Horizontal flip: Randomly flip the input image horizontally with $50\%$ using probability; \textbf{Level 3} + RandAugment \cite{cubuk2020randaugment}: Randomly apply rand augment twice with magnitude $9$ and magnitude standard deviation $0.5$; \textbf{Level 4} + Random Erasing \cite{zhong2020randomerasing}: Select a rectangle region of the input image and do pixel-wise erasing with $25\%$ using probability. The size of the selected region are randomly picked in the range $[0.02, 1/3]$ of the image size. \textbf{Level 5} + Mixup \cite{zhang2017mixup}: Mix two in-batch images up with a ratio $\lambda$ subjecting to a beta distribution, $\lambda \sim B(1, 1)$. The mixup process for images $I_A$ and $I_B$ can be formatted as $I_A(x,y)$ = $\lambda I_A(x,y) + (1-\lambda)I_B(x,y)$, where $(x,y)$ denotes the pixel coordinate; \textbf{Level 6} + CutMix \cite{yun2019cutmix}: Switch from Mixup to Cutmix with $50\%$ probability. Randomly replace a square region in the original image with a region in another in-batch image. The region size is randomly determined, subject to the squared root of a beta distribution $B(1, 1)$.
%

\begin{figure*}[pos=!ht]
    \centering
    \includegraphics[width=0.8\linewidth]{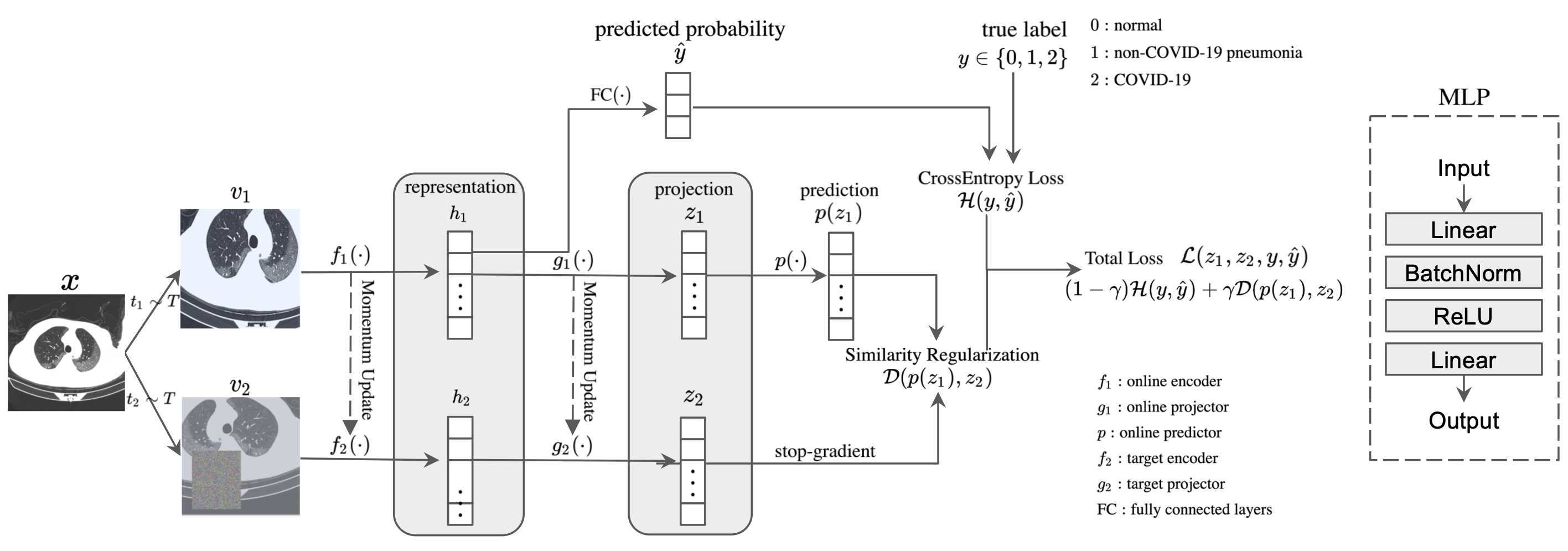}
    \caption{The overall structure of the models with our proposed similarity regularization in training interface. The two projectors, $g_1$ and $g_2$, and the online predictor $p$ are implemented by non-linear MLPs. After training, only the online encoder $f_1$ and fully connected layer $\text{FC}$ are preserved in testing.}
    \label{fig:structure}
\end{figure*}

The visualization of the augmented scans is demonstrated in Fig. \ref{fig:augmentation} in Appendix~\ref{sec:appendix_augmenation}. Specifically, RandomResizedCrop and horizontal flip are two commonly used augmentation operations in both supervised \cite{he2016resnet, szegedy2016inception} and self-supervised learning \cite{chen2020simclrv2, chen2020simclr, chen2020mocov2, grill2020byol}. Since contrastive learning requires more complicated augmentation \cite{chen2020simclr}, the two stronger augmentations, RandAugment and Random Erasing, are further introduced in levels 3 and 4. Their implementations and parameters refer to \cite{touvron2021training, rw2019timm}. In levels 5 and 6, mixup and cutmix are two augmentations enabling higher data diversity by fusing in-batch images. In these two levels, we mainly experiment on whether such sample-fusing augmentations can improve COVID-19 classification. By comparing the performance of models under these incremental augmentation levels, an appropriate augmentation strategy for COVID-19 CT scans can be established.

\subsection{Similarity Regularization}
\label{subsec:similarity_reg}
Most mainstream conventional CNNs contains two parts, a representation extractor $f$ and a followed fully connected layer $\text{FC}$. The extractor $f$ aims to extract the distinguishable representations of given inputs, and $\text{FC}$ predicts the class probability distribution by summarizing the extracted representations. This forward propagation is demonstrated as the top branch in Fig. \ref{fig:structure}. More formally, the input image $x$ is first transformed to a view $v$ by a random on-the-fly augmentation operation $t\sim T$ where $T$ denotes an infinite collection of augmentation operations. Subsequently, the representation extractor $f$ converts the input view $v$ to a representation embedding vector $h=f(v)$. The followed $\text{FC}$ predicts the class probability distribution based on its obtained representation, $\hat{y}=\text{FC}(h)$. The training target of such a classifier is to minimize the class probability distribution distance between the prediction $\hat{y}$ and the ground truth $y$ according to the cross-entropy loss in Eq. (\ref{eq:cross_entropy}), where $i \in \{0,1,2\}$ denotes the class index.
\begin{equation}
    \mathcal{H}(y,\hat{y}) = -\sum_{i=0}^2 y_i\log{\hat{y}_i}
    \label{eq:cross_entropy}
\end{equation}

In this conventional fully supervised scenario, the trained representations aim at better projecting to human-specified class distribution. However, this manner affects data efficiency, robustness or generalization \cite{khosla2020supervised}. Instead, contrastive learning enables learning more parameter-efficient representations from inputs themselves instead of the specified annotations. We thus incorporate it in common CNNs to improve their representation learning ability.

The overall structure of our method is illustrated in Fig. \ref{fig:structure}. We keep the conventional supervised classifier unchanged in the top branch while introducing a contrastive learning framework in the bottom branch. As in Section \ref{subsec:contrastive_learning}, contrastive learning is to maximize the representation similarity between positive pairs. We punish the positive-pair representation distance as a regularization term beside the cross-entropy loss, naming the term similarity regularization (SR). 

Particularly, the contrastive framework is a siamese network like most mainstream frameworks \cite{chen2020simclr, chen2020simclrv2, chen2021Simsiam, grill2020byol}, consisting of an online network and a target network. The target network can be seen as a moving average of the online one. Given two views $v_1$ and $v_2$ augmented from the same input image $x$, the representation extractors $f_1$ and $f_2$ in two networks extract their corresponding latent representation vectors, $h_1 = f_1(v_1)$ and $h_2=f_2(v_2)$. To avoid representations heavily affected by SR, the representation vectors then projected to another embedding space where $z_1=g_1(h_1)$ and $z_2=g_2(h_2)$, as in \cite{chen2020simclrv2, grill2020byol}. Since the projectors $g_1$ and $g_2$ share slightly different feature spaces, the online projection $z_1$ is further projected to $p(z_1)$ of same dimension via online predictor $p$. The cosine representation similarity $\mathcal{S}$ of value in range $[-1,1]$ can be measured according to Eq. (\ref{eq:cos_similarity}).

\begin{equation}
    \mathcal{S}(p(z_1), z_2) = \frac{\langle p(z_1), z_2 \rangle}{\lVert p(z_1) \rVert_2 \cdot \lVert z_2 \rVert_2 }
    \label{eq:cos_similarity}
\end{equation}
where $\langle\cdot, \cdot\rangle$ and $\lVert \cdot \rVert_2$ are inner product and $L^2$ norm notations, respectively. A higher value indicates two vectors are of higher similarity. To penalize a low cosine similarity between positive pairs and scale the penalty in range $[0,1]$, SR can be calculated as Eq. (\ref{eq:sr}).

\begin{equation}
\mathcal{D}(p(z_1), z_2) = 2 - 2 \mathcal{S}(p(z_1), z_2) = 2 - \frac{2\langle p(z_1), z_2 \rangle}{\lVert p(z_1) \rVert_2 \cdot \lVert z_2 \rVert_2 }
\label{eq:sr}
\end{equation}

Hence, for a positive pair $(v_1, v_2)$, its total loss containing both cross-entropy loss and SR is written as in Eq. (\ref{eq:total_loss}).
\begin{equation}
    \mathcal{L}(z_1, z_2, y, \hat{y}) = (1-\gamma)\mathcal{H}(y, \hat{y}) + \gamma\mathcal{D}(p(z_1), z_2)
    \label{eq:total_loss}
\end{equation}
where $\gamma$ is a scale factor for balancing the conventional cross-entropy loss and the introduced SR. 

$(v_2, v_1)$ is the symmetric positive pair with respect to $(v_1, v_2)$. We calculate the losses for both symmetric pairs and take their mean as the final loss for fast convergence.

SR as a regularization term may raise the concern if it will dominate the combined loss and thus degrade the classification. To remove the concern and find an appropriate scheduler for $\gamma$, we design three strategies as listed below. $i$ denotes the current training iteration number.
\begin{description}
\item[Constant (default)]: $\gamma$ is set to be a constant value during all training iterations, $0.5$ by default.

\item[Linear Decay]: $\gamma$ decays linearly to a minimum value $\gamma_{min}=0.01$ along $N$ training iterations according to Eq. (\ref{eq:gamma_linear}).
\begin{equation}
    \gamma_i = \gamma_{min} + (1-\frac{i}{N})(1-\gamma_{min})
    \label{eq:gamma_linear}
\end{equation}

\item[Cosine Decay]: $\gamma$ decays to a minimum value $\gamma_{min}=0.01$ along $N$ training iterations according to cosine annealing scheduler as in Eq. (\ref{eq:gamma_cosine}).
\begin{equation}
    \gamma_i = \gamma_{min} + \frac{1}{2}(1+\cos{\frac{i\pi}{N}})(1-\gamma_{min})
    \label{eq:gamma_cosine}
\end{equation}
\end{description}

\begin{algorithm}[!t]
\small
\caption{Algorithm for training models with our proposed similarity regularization.}
\label{algo:sr}
\SetKwInOut{Input}{Inputs}\SetKwInOut{Output}{output}
\SetKw{KwRet}{Return}
\Input{batch size $B$; iterations $N$; network components $f_1$, $f_2$, $g_1$, $g_2$, $p$, $\text{FC}$; cross entropy criterion $\mathcal{H}$; SR criterion $\mathcal{D}$; loss scale factor $\gamma$; momentum rate $\beta$. \\
}

\For{$i=1$ \KwTo $N$}{
    Sample a minibatch $\{(x_k, y_k)\}_{k=1}^{B}$\;
    \For{$k=1$ \KwTo $B$}{
        \tcp{Augmentation and predict representation}
        $t_1 \sim T$, $t_2 \sim T$\;
        $v_1=t_1(x_k)$, $v_2=t_2(x_k)$\;
        $h_1^{(1)}=f_1(v_1)$, $h_2^{(1)}=f_2(v_2)$\;
        $h_1^{(2)}=f_1(v_2)$, $h_2^{(2)}=f_2(v_1)$\;
        \tcp{Calculate cross-entropy}
        $\hat{y}^{(1)}=\text{FC}(h_1^{(1)})$, $\hat{y}^{(2)}=\text{FC}(h_1^{(2)})$\;
        $\mathcal{L}_k^{\text{CE}} = 0.5\mathcal{H}(y_k, \hat{y}^{(1)}) + 0.5\mathcal{H}(y_k, \hat{y}^{(2)})$\;
        \tcp{Calculate similarity regularization}
        $pz_1^{(1)}=p(g_1(h_1^{(1)}))$, $z_2^{(1)}=g_2(h_2^{(1)})$\;
        $pz_1^{(2)}=p(g_1(h_1^{(2)}))$, $z_2^{(2)}=g_2(h_2^{(2)})$\;
        Stop gradient for $z_2^{(1)}$ and $z_2^{(2)}$\;
        $\mathcal{L}_k^{\text{SR}} = 0.5\mathcal{D}(pz_1^{(1)}, z_2^{(1)}) + 0.5\mathcal{D}(pz_1^{(2)}, z_2^{(2)})$\;
    }
    \tcp{Scale the batch loss and update networks}
    Set $\gamma_i$ according to applied scheduler\;
    $\mathcal{L}_{\text{total}} = \frac{1}{B}\sum_{k=1}^{B}[(1-\gamma_i)\mathcal{L}_k^{\text{CE}} + \gamma_i \mathcal{L}_k^{\text{SR}}]$\;
    Update $f_1$, $g_1$, $p$, $\text{FC}$ by back-propagating $\mathcal{L}_{\text{total}}$\;
    \tcp{Momentum update weights of $f_2$ and $g_2$}
    $f_2$ = $(1-\beta)f_1 + \beta f_2$\;
    $g_2$ = $(1-\beta)g_1 + \beta g_2$\;
}
\KwRet{$f_1$ and $\text{FC}$;}
\end{algorithm}

Besides, after training, we throw away all the components except the online representation extractor $f_1$ and the fully connected layer $\text{FC}$. Hence, introducing SR in training will not slow down the test interface. The training pseudocode of models with SR is demonstrated in Algorithm \ref{algo:sr}.

%% file: sections/4_results.tex
\section{Results and Analysis}
\label{sec:results}

\subsection{Dataset Description}
\label{subsec:dataset}
In this paper, we mainly train and evaluate our proposed method using the largest existing open-access COVID-19 CT dataset, COVIDx CT-2A\footnote[2]{\url{https://www.kaggle.com/hgunraj/covidxct}}. Specifically, the dataset contrains three classes, including normal, non-COVID-19 pneumonia (NCP), and COVID-19 pneumonia (CP). Its class distribution is summarized in Table~\ref{tab:class_dist}. The dataset is of high diversity, containing scans of $3,745$ patients from eight open-access sources. It should be noted that the scans from the same patient are in one subset, preventing information leakage from training to validation or testing.
\begin{table}[pos=!t]
    \centering
    \caption{Class distribution of the employed COVIDx CT-2A dataset.}
    \label{tab:class_dist}
    \begin{tabular*}{\columnwidth}{@{\extracolsep{\fill}}lcccc@{}}
    \toprule
    Set         &   Normal      &   NCP         &   CP       &   Total       \\
    \midrule
    Training    &   $35,996$    &   $25,496$   &   $82,286$   &   $143,778$      \\
    Validation  &   $11,842$    &   $7,400$    &   $6,244$    &   $25,486$       \\
    Testing     &   $12,245$    &   $7,395$    &   $6,018$    &   $25,658$       \\
    \bottomrule
    \end{tabular*}
\end{table}

\subsection{Experimental Setting}
\label{subsec:implementation}
In this paper, we keep the hyper-parameters consistent in all experiments for fair comparisons. The codes are implemented by PyTorch. We implement the CNN backbones and image augmentation by torchvision and timm \cite{rw2019timm} libraries, respectively. For acceleration, we train models on Torch distributed data parallelism on four Nvidia V100 GPUs with apex mixed precision of level $O1$. Besides, to alleviate the randomness concern, we obtain the experimental statistics by averaging the measurements in three distinct trials.

During training, CT scans are resized to $256\times256$ in $3$ channels using bicubic interpolation and normalized by ImageNet mean and std. In the test interface, $256\times256$ CT scans are cropped from the center of resized $293\times293$ original images. This is empirically good as the center crop can preserve the main lung regions. To avoid models from being too confident in one-class prediction, label smoothing \cite{szegedy2016inception} of smoothing factor $0.1$ is applied in the  cross-entropy in augmentation levels $0-4$. While in augmentation level $5$ or $6$, in-batch paired labels are mixed up based on mixed inputs (See \cite{zhang2017mixup, yun2019cutmix} for more details). 

The optimizer we used is Adam with $10^{-6}$ weight decay. After a 5-epoch linear warmup \cite{loshchilov2016sgdr} from $5\times10^{-7}$, we use cosine annealing scheduler to decay the learning rate from $5\times10^{-4}$ to $5\times10^{-7}$ in the later $45$ epochs. The batch size is set to be $64$ in each process. Besides, the gradients are clipped to be no larger than $5.0$ to avoid overflow. 

In the SR calculation, the projectors $g_1, g_2$ and predictor $p$ have the same multi-layer perceptron (MLP) architecture that consists of two linear layers connected by a batch normalization layer and a ReLU activation layer. The front linear layer projects the inputs to $512$-D embedding vectors and the later linear layer outputs $128$-D vectors. The analysis for the dimension setting is in Appendix~\ref{sec:appendix_projection}.
The momentum rate $\beta$ for updating $f_2$ and $g_2$ is $0.99$, a median value among contrastive frameworks \cite{he2020mocov1, chen2021Simsiam, grill2020byol, caron2020swav}. 

\subsection{Results of ResNets under Incremental Augmentation Levels}
\label{subsec:acc_resnet}
We first compare the performance of ResNets with or without SR under the incremental augmentation levels designed in Section \ref{subsec:augmentation} to determine an appropriate augmentation policy for the coming experiments. The averaged test accuracies are listed in Table~\ref{tab:acc_resnets}. Since SR requires calculating the similarity between two augmented views, models with SR cannot be implemented under augmentation level $0$.

\begin{table}[pos=!ht] \centering
    \caption{Test accuracy comparison between original ResNets and ResNets with proposed SR under incremental augmentation levels. ResNet is abbreviated as R. The scale factor scheduler for scaling SR is the default constant scheduler $\gamma=0.5$. }
    \label{tab:acc_resnets}
    \begin{tabular*}{\columnwidth}{@{\extracolsep{\fill}}ccccc@{}}
    \toprule
    \multirow{2}{*}{Augmentation}    &   \multirow{2}{*}{Method}    &   \multicolumn{3}{c}{CNN Backbone}    \\ \cmidrule{3-5}
                                &               &   R18    &   R34    &   R50    \\ \midrule
    Level $0$                   &   Original    &   $91.28$     &   $92.28$     &   $91.57$     \\ \midrule
    \multirow{2}{*}{Level $1$}  &   Original    &   $99.10$     &   $99.00$     &   $98.89$     \\
                                &   +SR(Ours)   &   $99.23$     &   $99.27$     &   $99.09$     \\ \midrule
    \multirow{2}{*}{Level $2$}  &   Original    &   $99.17$     &   $99.11$     &   $99.00$     \\
                                &   +SR(Ours)   &   $99.27$     &   $99.39$     &   $99.19$     \\ \midrule
    \multirow{2}{*}{Level $3$}  &   Original    &   $99.11$     &   $99.06$     &   $99.08$     \\
                                &   +SR(Ours)   &   $99.26$     &   $99.29$     &   $99.20$     \\ \midrule
    \multirow{2}{*}{Level $4$}  &   Original    &   $99.12$     &   $99.09$     &   $99.13$     \\
            &   +SR(Ours)   &   $\mathbf{99.40}$     &   $\mathbf{99.43}$     &   $\mathbf{99.31}$     \\ \midrule
    Level $5$                   &   Original    &   $97.49$     &   $97.34$     &   $97.66$     \\ \midrule
    Level $6$                   &   Original    &   $98.50$     &   $98.69$     &   $98.91$     \\
    \bottomrule
    \end{tabular*}
\end{table}

\begin{table*}[!ht]\centering
    \caption{Test accuracy of seven CNNs with or without SR under augmentation level $4$, ordered by the number of parameters. The scale factor scheduler for scaling SR is the default constant strategy $\gamma=0.5$. Note that the extra parameters introduced by SR will be discarded in test interface after training.}
    \label{tab:acc_all_models}
    \begin{tabular*}{\linewidth}{@{\extracolsep{\fill}}ccccccccc@{}}
    \toprule
    \multirow{2}{*}{Method} & \multirow{2}{*}{Metric} & \multicolumn{7}{c}{CNN Backbone}    \\ \cmidrule{3-9}
        &   &   SqueezeNet & MobileNet & DenseNet  &  ResNet18 &   ResNet34  & InceptionV3 &  ResNet50 \\ \midrule
    \multirow{2}{*}{Original}   & Params. (M) & $0.69$   & $2.12$  & $6.63$  & $10.66$  & $20.30$ & $20.78$ & $22.42$ \\
                                & Acc. (\%)   & $98.22$  & $99.02$ & $99.05$ & $99.08$  & $99.06$ & $99.22$ & $99.12$ \\ \midrule
    \multirow{2}{*}{+SR(Ours)}  & Params. (M) & $1.13$   & $2.94$  & $7.33$  & $11.10$  & $20.74$ & $21.97$ & $23.62$ \\
                                & Acc. (\%)   & $98.41$  & $99.18$ & $\mathbf{99.44}$ & $99.39$  & $99.43$ & $99.32$ & $99.31$ \\
    \bottomrule
    \end{tabular*}
\end{table*}

\begin{table*}[pos=!ht] \centering
    \caption{Comparison of DenseNet121-SR with the state-of-the-art methods on COVIDx CT-2A dataset.}
    \label{tab:measurements}
    \begin{tabular*}{\linewidth}{@{\extracolsep{\fill}}lcccccccccc@{}}
    \toprule
    \multirow{2}{*}{Method} & \multirow{2}{*}{Accuracy (\%)} & \multicolumn{3}{c}{Precision (\%)} & \multicolumn{3}{c}{Sensitivity (\%)} & \multicolumn{3}{c}{Specificity (\%)} \\
    \cmidrule(lr){3-5} \cmidrule(l){6-8} \cmidrule(l){9-11}
    & & Normal & NCP & CP & Normal & NCP & CP & Normal & NCP & CP \\ \midrule
    COVID-Net CT-1 \cite{Gunraj2020ct1} & $94.5$ & $96.1$ & $97.6$ & $90.2$ & $98.8$ & $80.2$ & $99.0$ & $96.3$ & $99.4$ & $95.7$ \\
    COVID-Net CT-2 S \cite{Gunraj2021ct2} & $97.9$ & $99.3$ & $96.4$ & $97.0$ & $98.9$ & $95.7$ & $98.1$ & $99.3$ & $98.9$ & $98.8$ \\
    COVID-Net CT-2 L \cite{Gunraj2021ct2} & $98.1$ & $99.4$ & $96.7$ & $97.2$ & $99.0$ & $96.2$ & $98.2$ & $99.5$ & $99.0$ & $98.8$ \\
    Bit-M \cite{zhao2021deep} & $99.2$ & $99.8$ & $98.9$ & $\mathbf{98.5}$ & $\mathbf{99.3}$ & $99.6$ & $98.7$ & $99.8$ & $99.6$ & $99.5$ \\
    DenseNet121-SR (Ours) & $\mathbf{99.44}$ & $\mathbf{99.89}$ & $\mathbf{99.55}$ & $98.40$ & $99.12$ & $\mathbf{99.83}$ & $\mathbf{99.59}$ & $\mathbf{99.91}$ & $\mathbf{99.82}$ & $\mathbf{99.50}$ \\ \bottomrule
    \end{tabular*}
\end{table*}

Table~\ref{tab:acc_resnets} shows that the original ResNets achieve the highest accuracy in level $2$ and cannot be improved in the following levels, heavily degraded in levels $5$ and $6$. The degradation may result from the fact that sample-fusing augmentation sometimes transfers the pneumonic pathologies from CP/NCP cases to normal cases. We thus do not perform SR in levels $5$ and $6$. Different from the original ResNets, ResNets with SR continue to improve after level $2$ and achieve the highest accuracy in level $4$. This is consistent with the findings in many contrastive learning research works that contrastive learning requires stronger augmentation than supervised models \cite{chen2020simclr, chen2020mocov2, asano2019critical}. Hence, we select level $4$ as the basic augmentation level for the following experiments. Overall, it is observed that SR can improve the classification performance of ResNets stably under all augmentation levels from $1$ to $4$.

\subsection{Results under Augmentation Level 4}
\label{subsec:acc_level_4}
In this section, we extend the experiments to seven widely used CNNs, including SqueezeNet1.1 \cite{iandola2016squeezenet}, MobileNetV2 \cite{sandler2018mobilenetv2}, DenseNet121 \cite{huang2017densenet}, ResNet-18/34/50 \cite{he2016resnet}, and InceptionV3 \cite{szegedy2016inception}. The experiments are under augmentation level 4 and a constant scale factor scheduler $\gamma=0.5$.

From the results in Table~\ref{tab:acc_all_models}, it can be seen that all our models with SR surpass the original models in terms of averaged test accuracy. The best model, DenseNet121 with SR, achieves $99.44\%$ accuracy with $7.33M$ parameters. Note that the extra parameters will be thrown away after training so that the parameters in the test interface are consistent for a model with or without SR. Another observation is that, in COVID-19 CT classification, the model performance is not in strictly proportional to its capacity despite of model architecture. This suggests that the fine design of model architecture rather than simply expanding depth or width is more valuable in this task, as supported in \cite{Gunraj2020ct1, Gunraj2021ct2, polsinelli2020light}. 

\begin{figure}[!ht]
    \centering
    \includegraphics[width=1.0\columnwidth]{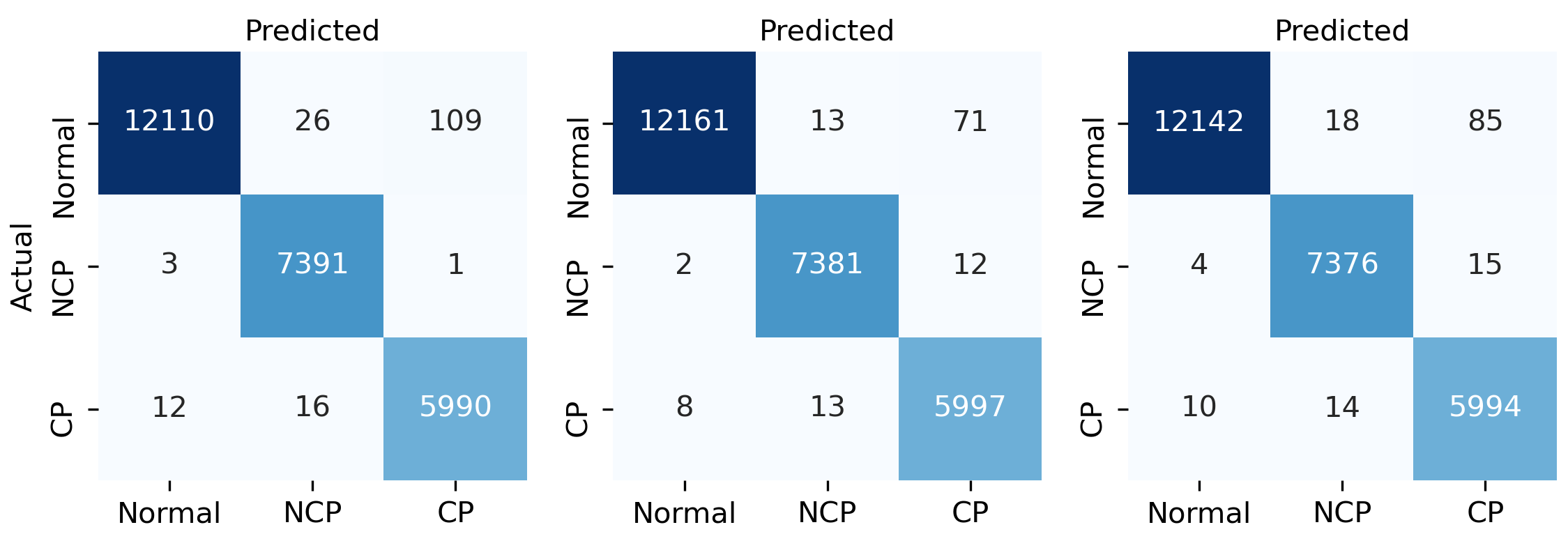}
    \caption{Confusion matrix for DenseNet121-SR in three trials.}
    \label{fig:cm_densen121_sr}
\end{figure}

Fig. \ref{fig:cm_densen121_sr} shows the confusion matrices for DenseNet121-SR in three training trials. Based on the matrices, we measure the performance of the model in terms of averaged accuracy, precision, sensitivity, and specificity, as listed in Table~\ref{tab:measurements}. The results show that our DenseNet121-SR has outperformed the state-of-the-art models in nearly all measurements. Specifically, DenseNet121-SR achieves a high sensitivity $99.59\%$ for COVID-19 positive class, indicating that the model has the potential to efficiently avoid COVID-19 positive patients from being wrongly diagnosed. 

Besides, to better understand the classification principles of the model, its attention visualized by Grad-CAM \cite{selvaraju2017grad} is demonstrated in Fig. \ref{fig:attention} in Appendix~\ref{sec:appendix_visualization}.

\subsection{Results on Other Datasets}
\label{subsec:other_datasets}
\textbf{On other COVID-19 CT datasets}
Based on the experimental results aforementioned, we extend our method to the other two COVID-19 CT datasets, i.e., SARS-CoV2 and COVIDx-CT-1. It should be noted that, for SARS-CoV2, we train DenseNet121-SR over $200$ epochs with weights pre-trained on ImageNet because SARS-CoV2 contains much fewer CT scans than the others. The results as listed in Table~\ref{tab:result_sarscov2_covidxct1} show that our method can be generalized to other datasets and can achieve a high classification performance. Comparing to the methods listed in Table~\ref{tab:comparison_sota}, our DenseNet121-SR with only $6.63$ MB parameters is more parameter-efficient and outperforms the reviewed methods.

\begin{table}[pos=!ht]
    \centering
    \caption{Classification results of DenseNet121-SR on SARS-CoV2 and Covidx CT-1 datasets. The precision, sensitivity, and specificity metrics are for COVID-19 positive class only.}
    \label{tab:result_sarscov2_covidxct1}
    \begin{tabular*}{\linewidth}{@{\extracolsep{\fill}}lcccc@{}}
    \toprule
    Dataset     & Acc (\%) & Prec (\%) & Sens (\%) & Spec (\%) \\ \midrule
    SARS-CoV2   & $99.20$  & $99.47$   & $98.93$     & $99.46$     \\
    COVIDx CT-1 & $99.78$  & $99.56$   & $99.84$     & $99.74$      \\
    \bottomrule
    \end{tabular*}
\end{table}

\textbf{On classic natural datasets}
Besides, extensive experiments are conducted over seven natural datasets to further evaluate the generalization ability of our method. To evaluate the effect of SR fairly, we keep the setting unchanged as in Section \ref{subsec:implementation} and initialize the model weights as pre-trained on ImageNet. Table \ref{tab:result_classic_datasets} demonstrates the classification accuracy of DenseNet121 with or without SR on the seven datasets, including FGVC Aircraft \cite{fgvcaircraft}, CIFAR10/100 \cite{cifar}, Describable Textures Dataset (DTD) \cite{dtd}, Oxford 102 Flowers \cite{flowers102}, Oxford-IIIT Pets \cite{oxfordpets}, and Stanford Cars \cite{stanfordcars}. It shows that DenseNet121-SR is superior to the original model in all the tasks, indicating our proposed SR can be generalized to general classification problems.

\begin{table*}[pos=!ht]
    \centering
    \caption{Classification accuracy of DenseNet121 with or without SR over seven classic natural datasets.}
    \label{tab:result_classic_datasets}
    \begin{tabular*}{\linewidth}{@{\extracolsep{\fill}}lccccccc@{}}
    \toprule
                        & Aircraft & CIFAR10 & CIFAR100 & DTD     & Flowers102 & OxfordIIITPet & StanfordCars \\ \midrule
        DenseNet121     & $88.15$  & $94.45$ & $85.08$  & $70.60$ & $93.17$    & $92.47$       & $92.46$      \\
        DenseNet121-SR (Ours)  & $\mathbf{88.18}$ & $\mathbf{94.47}$ & $\mathbf{85.08}$ & $\mathbf{71.01}$ & $\mathbf{94.42}$ & $\mathbf{92.88}$ & $\mathbf{92.55}$ \\
    \bottomrule
    \end{tabular*}
\end{table*}

%% file: sections/5_ablation.tex
\section{Ablation Study}
\label{sec:ablation}
The following ablation studies are conducted to better investigate the effects of our proposed SR. 


\textbf{Fully self-supervised learning} Contrastive learning is widely adopted in pre-training CNNs that are fine-tuned later for downstream tasks. In our methods, we turn the process to an end-to-end manner by regularizing CNNs with proposed SR derived from contrastive learning. Hence, a comparison between SR and conventional contrastive learning is necessary. Specifically, we design and measure the following methods for comparison as follows. 
\begin{enumerate}[a)]
    \item Linear Evaluation. First pre-train the representation extractor $f$ weights of which are frozen in the later $\text{FC}$ fine-tuning. The pre-training process is equivalent to setting $\gamma=1$ in all training epochs as in Algorithm \ref{algo:sr}, and then fine-tuning only the linear layers $\text{FC}$ as usual. The fine-tuning hyper-parameters include: $256$ batch size, learning rate decays from $40$ to $4\times10^{-6}$ according to cosine decay scheduler \cite{loshchilov2016sgdr}. The optimizer used is SGD. Linear evaluation is simply conducted for verifying the effects of contrastive learning in this task.
    \item Two-stage training, self-supervised contrastive learning followed by conventional supervised learning. Pre-training the representation extractor $f$ and training the entire CNN with pre-trained weights. The hyper-parameters are consistent with others as in Section \ref{subsec:implementation}.
    \item Apply SR to ResNets with a default constant $\gamma=0.5$. 
\end{enumerate}

\begin{table}[!ht]   \centering
    \caption{Test accuracy comparison between two-stage training and end-to-end training with SR under incremental augmentation levels. The scale factor scheduler for scaling SR is the default constant scheduler $\gamma=0.5$.}
    \label{tab:abalation_finetunne}
    \begin{tabular*}{\columnwidth}{@{\extracolsep{\fill}}ccccc@{}}
    \toprule
    \multirow{2}{*}{Augmentation}    &   \multirow{2}{*}{Method}    &   \multicolumn{3}{c}{CNN Backbone}    \\ \cmidrule{3-5}
                                &             &   R18   &   R34   &   R50    \\ \midrule
    \multirow{3}{*}{Level $1$}  & Linear Eval & $95.47$ & $92.89$ & $92.32$  \\
                                & Two-stage   & $99.17$ & $99.17$ & $99.04$  \\
                                & +SR(Ours)   & $\mathbf{99.23}$ & $\mathbf{99.27}$ & $\mathbf{99.09}$  \\ \midrule
    \multirow{3}{*}{Level $2$}  & Linear Eval & $95.76$ & $92.03$ & $94.02$  \\
                                & Two-stage   & $99.25$ & $99.23$ & $\mathbf{99.20}$  \\
                                & +SR(Ours)   & $\mathbf{99.27}$ & $\mathbf{99.39}$ & $99.19$  \\ \midrule
    \multirow{3}{*}{Level $3$}  & Linear Eval & $93.92$ & $93.88$ & $95.08$  \\
                                & Two-stage   & $99.26$ & $99.29$ & $\mathbf{99.24}$  \\
                                & +SR(Ours)   & $\mathbf{99.26}$ & $\mathbf{99.29}$ & $99.20$  \\ \midrule
    \multirow{3}{*}{Level $4$}  & Linear Eval & $93.12$ & $93.23$ & $94.53$  \\
                                & Two-stage   & $99.38$ & $99.30$ & $99.18$  \\
                                & +SR(Ours)   & $\mathbf{99.40}$ & $\mathbf{99.43}$ & $\mathbf{99.31}$ \\
    \bottomrule
    \end{tabular*}
\end{table}

The results for the designs above are listed in Table~\ref{tab:abalation_finetunne}. It can be observed that contrastive learning can learn efficient representations that even a simple linear evaluation on the pre-trained representation extractor can achieve over $92\%$ test accuracy. For the second method, two-stage contrastive learning, the pre-trained weights from the representation extractor might be hard to maintain in the later training phase. Our introduced SR maintains the representation by explicitly penalizing the representation difference for positive pairs. The results in Table~\ref{tab:abalation_finetunne} verify that ResNets with SR surpass the two-stage contrastive learning method in most experiments. Besides, it is worth noting that the end-to-end training with SR does not require pre-training and thus saves computational resources.

\begin{figure}[!ht]
    \centering
    \includegraphics[width=0.65\columnwidth]{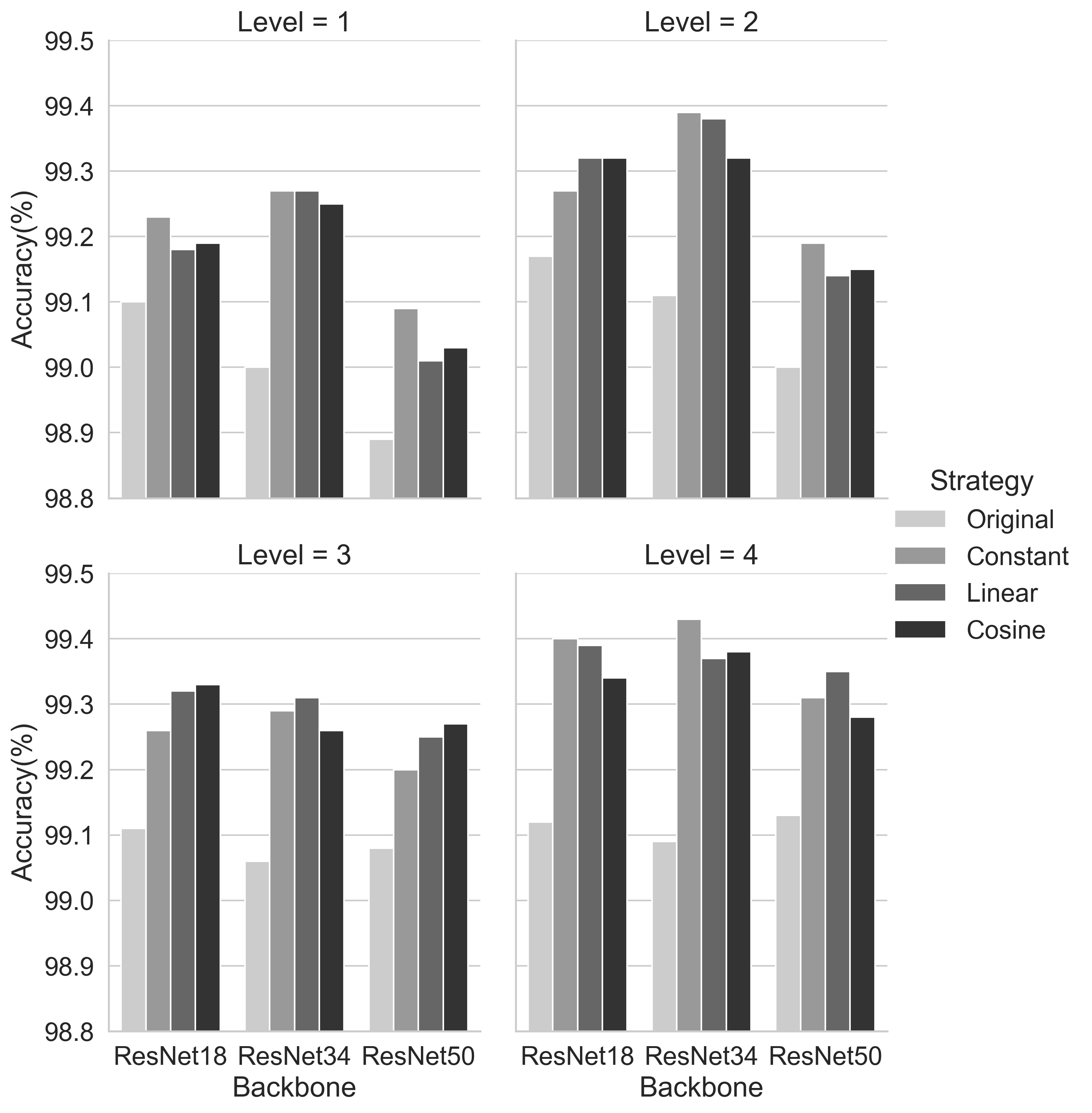}
    \caption{Accuracy of ResNets obtained with different $\gamma$ decay strategies under incremental augmentation levels from 1 to 4. The baselines are the original models without introduced SR. The $\gamma$ value for the constant strategy is $0.5$ by default while $\gamma$ decays from $1.0$ to $0.01$ in linear and cosine strategies.}
    \label{fig:gamma_strategies}
\end{figure}

\textbf{Decay strategy for $\gamma$} The two-stage contrastive learning method can be approximate to run the SR Algorithm \ref{algo:sr} with $\gamma=1$ in pre-training and $\gamma=0$ in fine-tuning. The sharp fall of $\gamma$ may destroy the maintained representation space obtained in contrastive pre-training. To avoid the potential negative impact, we designed two mild $\gamma$ decay strategies in Section \ref{subsec:similarity_reg} despite of the constant $\gamma$ strategy. From the results demonstrated in Fig. \ref{fig:gamma_strategies}, we can conclude that SR with all designed $\gamma$ strategies can stably improve the classification accuracy. And SR is insensitive to the $\gamma$ strategy setting since all strategies have comparable performance. Due to the simplicity of the $\gamma$ strategy ($\gamma=0.5$ in all iterations) and its slight superiority in level $4$ augmentation, we select it as the default strategy in our experiments. 

\textbf{$\gamma$ value in constant strategy} The ablation studies find that a constant strategy, $\gamma=0.5$, can achieve the relatively highest performance among the three strategies under level $4$ augmentation. We vary the constant $\gamma$ value to evaluate its robustness. See Appendix~\ref{sec:appendix_gamma_constant} for the detailed results.

%% file: sections/6_discussion.tex
\section{Discussion}
\label{sec:discussion}
Since COVID-19 grows rapidly worldwide, designing efficient and accurate classification systems is essential. Although some methods \cite{alshazly2021explainable, silva2020covid, Gunraj2020ct1, Gunraj2021ct2, zhao2021deep} have claimed a high classification accuracy ($\approx 99\%$) on multiple datasets, we argue that even a slight improvement can mitigate further infection. Meanwhile, some high-performance methods require considerable computational resources, making them hard to be deployed into practical healthcare systems. Hence, designing more efficient models with an affordable number of training parameters should also be noted. 

In this paper, we propose an incremental augmentation strategy and SR to improve the CNN classification performance on three COVID-19 CT datasets. The results illustrate that appropriate augmentation can significantly alleviate the data limitation problem in COVID-19 CT classification. Meanwhile, our proposed SR further improves the classification performance of seven CNNs by enhancing their representation learning ability. Specifically, on the largest dataset COVIDx CT-2A, our model DenseNet121-SR achieves $99.44\%$ accuracy and $99.59\%$ sensitivity with only $6.63$ MB parameters in the test interface, outperforming all the reviewed state-of-the-art methods. Besides, we evaluate DenseNet121-SR on the other two datasets, achieving $99.78\%$ and $99.20\%$ accuracy on COVIDx CT-1 and SARS-CoV2, respectively. To further justify the effect of SR, we extend the DenseNet121-SR to seven classic natural datasets, illustrating SR can be generalized to general classification tasks. Furthermore, since SR derives from contrastive learning, we compare traditional contrastive learning and end-to-end training with SR in ablation studies. The comparison demonstrates that SR is superior in classification accuracy and training efficiency and is robust to its hyper-parameter setting.

Despite the achieved promising performance, the limitations of our method exist. Our method requires either large amounts of training data or pre-training on other large-scale datasets. The high performance of our models with SR partly owes to the efforts of workers collecting numerous CT scans. For smaller-scale datasets like SARS-CoV2, the backbone of our method requires pre-training. The pre-training on ImageNet helps improve the accuracy from around $98\%$ to $99.20\%$ in the DenseNet121-SR case. Besides, due to the lack of computational resources, we can hardly evaluate other contrastive frameworks. Meanwhile, we cannot redesign the CNN backbones to better balance computational efficiency and classification performance because of the substantial computational loads of neural architecture searching and pre-training. For future work, we will explore redesigning the network backbone, pre-training the redesigned backbones on large-scale datasets, and making networks more explainable in COVID-19 CT diagnosis.

%% file: sections/7_conclusion.tex
\section{Conclusion}
\label{sec:conclusion}
This paper aims to improve the CNN performance for COVID-19 CT classification by enabling CNNs to learn parameter-efficient representations from CT scans. We propose the SR technique derived from contrastive learning and apply it to seven commonly used CNNs. The experimental results show that SR can stably improve the CNN classification performance. Together with a well-designed augmentation strategy, our model DenseNet121-SR with $6.63$ MB parameters outperforms the existing methods on three COVID-19 CT datasets, including SARS-CoV2, COVIDx CT-1, and COVIDx CT-2A. Specifically, on the largest available dataset COVIDx CT-2A, DenseNet121-SR achieves $99.44\%$ accuracy, with $98.40\%$ precision, $99.59\%$ sensitivity, and $99.50\%$ specificity for the COVID-19 pneumonia category. Furthermore, the extensive experiments on seven classic natural datasets demonstrate that SR can be generalized to common classification problems.

%% file: sections/acknowledgement.tex
\section*{Acknowledgement}
This work was supported by King's College London and has been performed using resources from the Cirrus UK National Tier-2 HPC Service at EPCC funded by the University of Edinburgh and EPSRC (EP/P020267/1).

%% file: sections/appendix.tex
\appendix
\section{Augmentation Visualization}
\label{sec:appendix_augmenation}
\begin{figure}[!ht]
    \centering
    \includegraphics[width=\columnwidth]{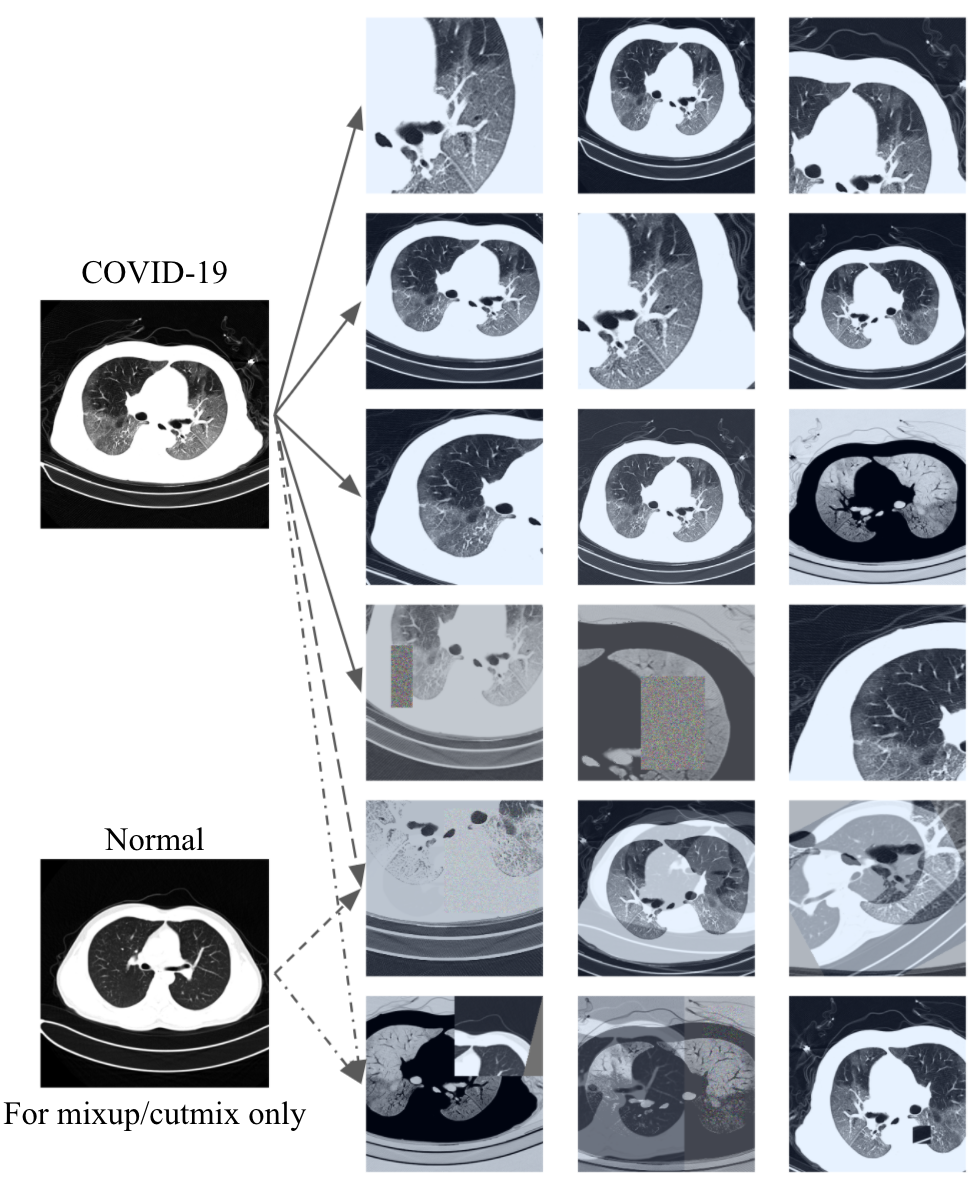}
    \caption{Illustration of applied incremental augmentation of six levels. On the right side, six groups of images are augmented from the same left COVID-19 positive scan. These groups from top to down are in the augmentation levels from 1 to 6. The bottom left normal scan is the auxiliary original image that only participated in Mixup/Cutmix augmentation in levels 5 and 6. The left two scans are from COVIDx CT-2A.}
    \label{fig:augmentation}
\end{figure}

The augmented CT scans under designed augmentation level from 1 to 6 are visualized in Fig. \ref{fig:augmentation}. These augmentation operations keep the main pneumonic pathologies reserved while enhancing the dataset diversity.

\section{Effect of Projection Size}
\label{sec:appendix_projection}
\begin{figure}[!ht]
    \centering
    \includegraphics[width=\columnwidth]{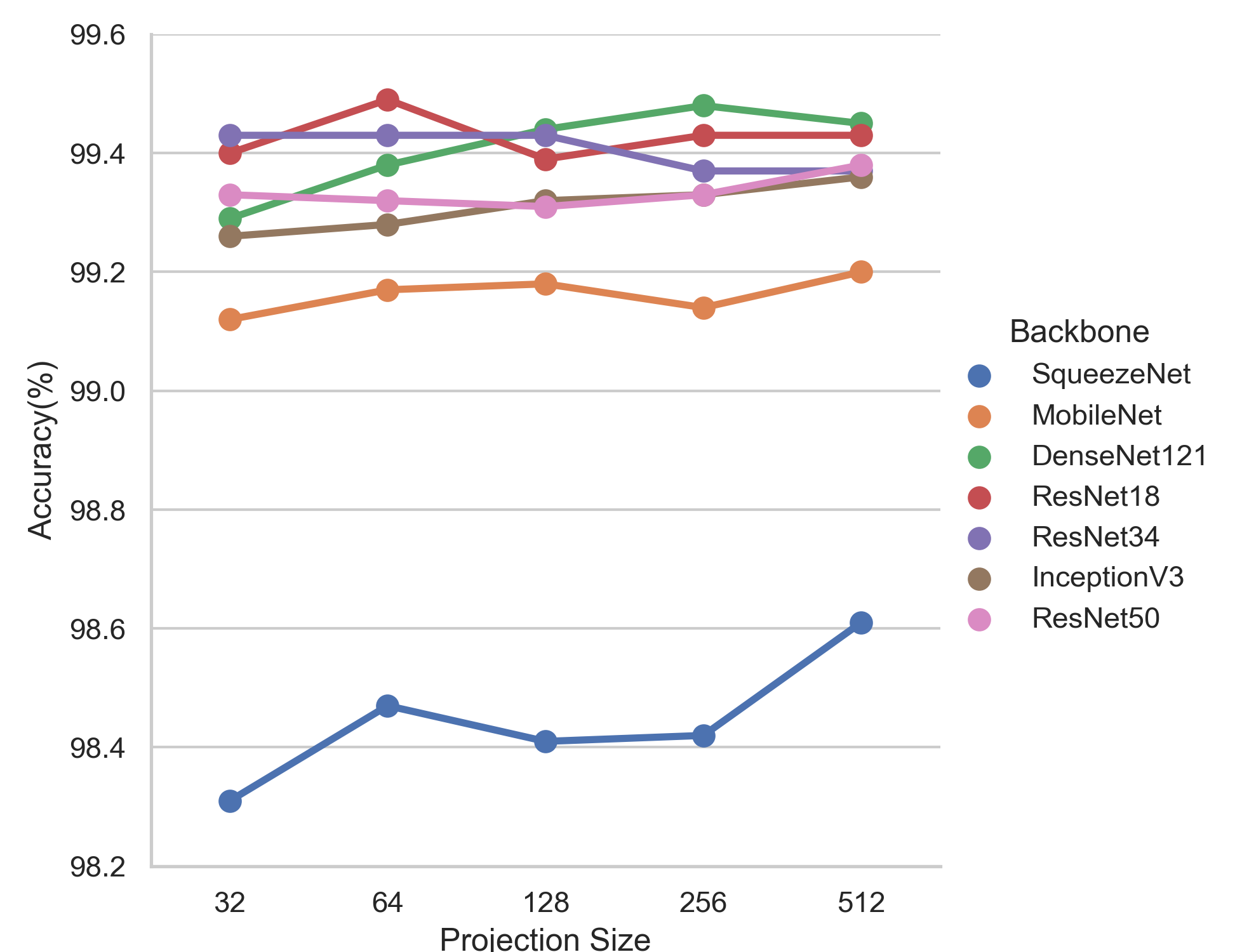}
    \caption{Impact of projection size in SR calculation.}
    \label{fig:projection}
\end{figure}
The output dimension, or named projection size, of both the projector and predictor in SR calculation are set to be $128$ as default. We keep the hidden dimension $512$ unchanged to avoid redundant computation while varying the projection size to analyze its effect in terms of classification accuracy. As visualized in Fig. \ref{fig:projection}, the differences between the classification accuracy for models except for SqueezeNet are small ($\leq 0.2\%$). This indicates that the hyper-parameter setting in our proposed SR is robust. 

\section{Attention Visualization}
\label{sec:appendix_visualization}
\begin{figure*}[!ht]
    \centering
    \includegraphics[width=\linewidth]{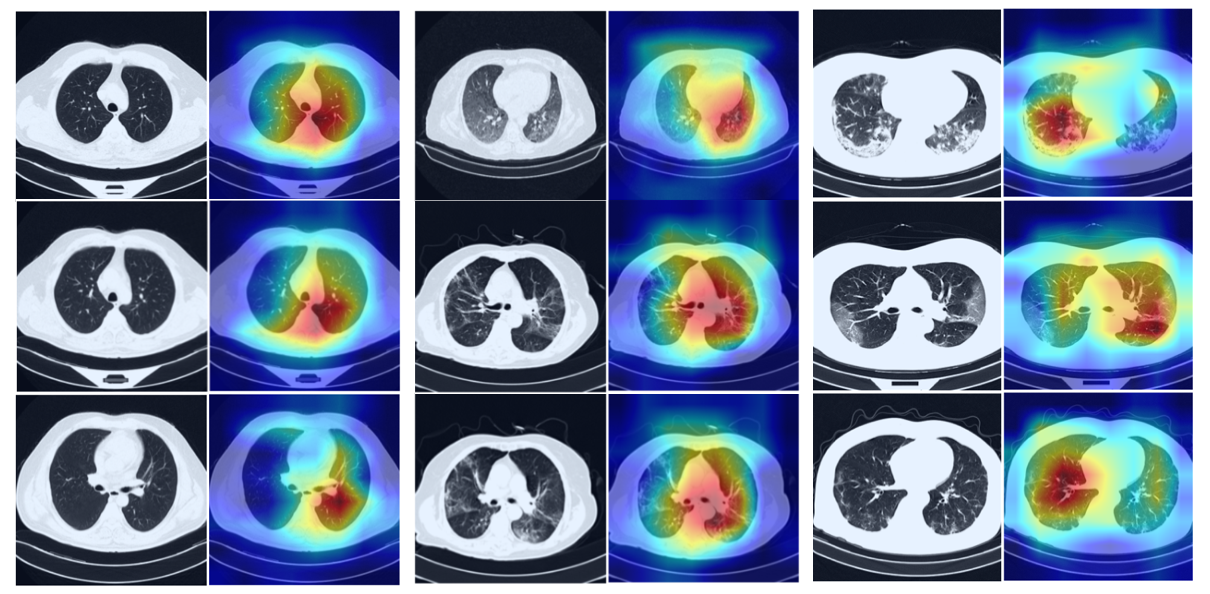}
    \caption{Attention of DenseNet121-SR visualized by Grad-CAM. The three groups of CT scans and heatmaps from left to right are in class normal, NCP, and CP, respectively. The highlighted parts are the regions based on which CNNs classify the CT scans.}
    \label{fig:attention}
\end{figure*}

To understand the behavior of our model, we visualize the attention of DenseNet121-SR on three CT scans in different classes as in Fig. \ref{fig:attention}. It can be observed that our model mainly focuses its attention on some suspicious regions where the pathologies may exist. 

\section{Effect of Constant $\gamma$ Strategy}
\label{sec:appendix_gamma_constant}
\begin{figure*}[!ht]
    \centering
    \includegraphics[width=\linewidth]{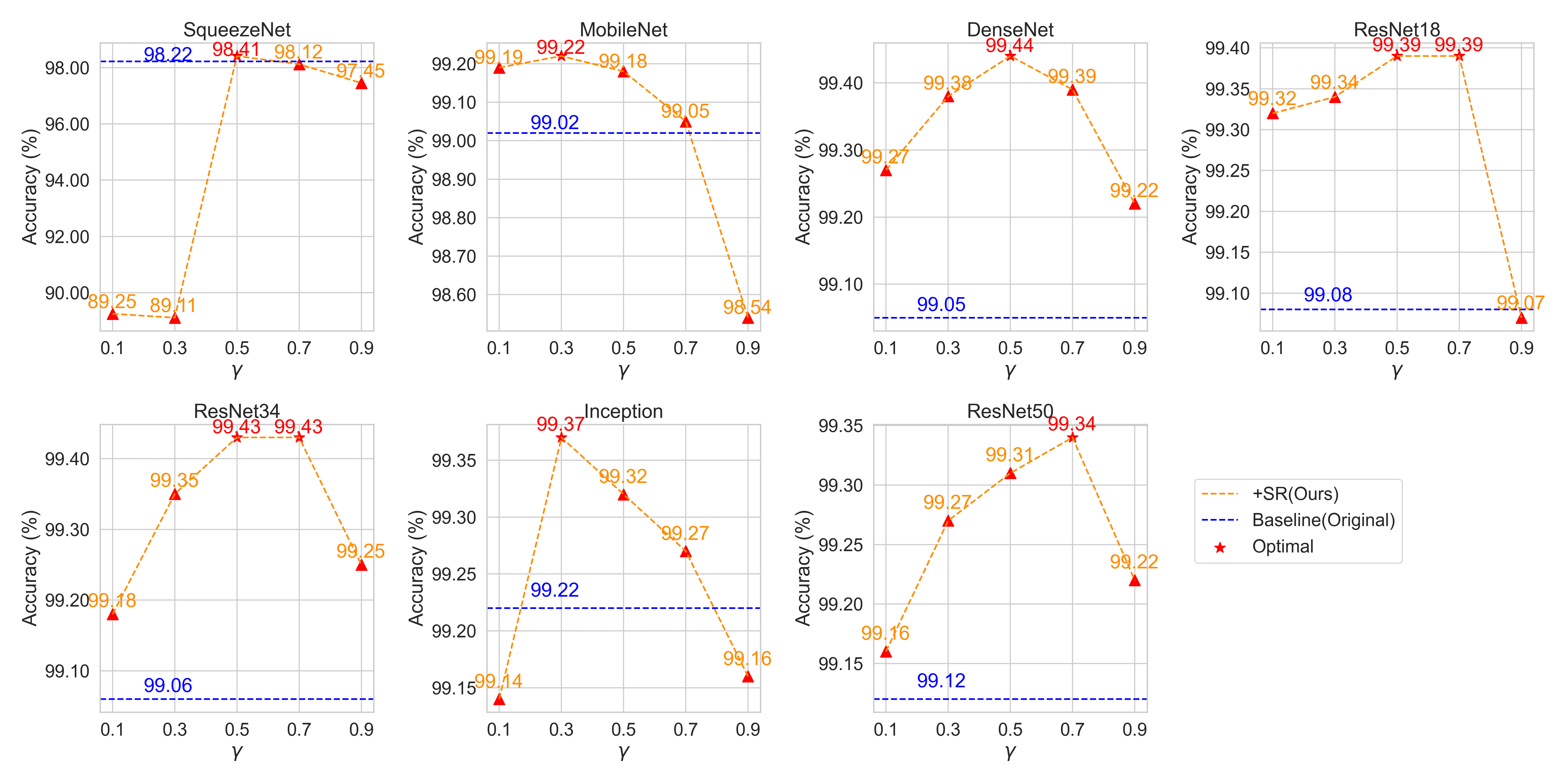}
    \caption{Accuracy of seven CNNs with SR controlled by constant $\gamma$ scale factor strategy, under augmentation level $4$. $\gamma$ varies from $0.1$ to $0.9$ with $0.2$ interval.}
    \label{fig:gamma_constant}
\end{figure*}

The constant $\gamma$ value still requires studies for finding its effects on model performance. We thus vary the $\gamma$ value in constant strategy from $0.1$ to $0.9$ with $0.2$ interval and repeat the experiments for CNNs with SR under augmentation level $4$. As shown in Fig. \ref{fig:gamma_constant}, SR can improve the CNN classification performance when $\gamma$ value is in an appropriate range near $[0.5, 0.7]$. In particular, a smaller $\gamma$ cannot fully fulfill the advantage of SR and sometimes even degrades the model capacity as in SqueezeNet1.1 case. Meanwhile, setting $\gamma$ to a large value like $0.9$ is also risky since SR dominates the total loss while the primary cross entropy for classification is slighted. 

\FloatBarrier